\documentclass[aps,prl,twocolumn,groupedaddress]{revtex4}  
\usepackage{graphicx}
\usepackage{latexsym}

\def \refeq#1{(\ref{#1})}
\def \reffig#1{Fig.\,\ref{#1}}
\def \reftab#1{Table\,\ref{#1}}

\begin{document}


\title{Explanation of the discrepancy between the measured and atomistically calculated yield stresses in
body-centered cubic metals}

\author{R. Gr\"{o}ger}
\email[]{groger@seas.upenn.edu}

\author{V. Vitek}
\affiliation{University of Pennsylvania, Department of Materials
  Science and Engineering, 3231 Walnut Street, Philadelphia, PA 19104}


\begin{abstract}
  We propose a mesoscopic model that explains the factor of two to three discrepancy between
  experimentally measured yield stresses of BCC metals at low temperatures and typical Peierls
  stresses determined by atomistic simulations of isolated screw dislocations. The model involves a
  Frank-Read type source emitting dislocations that become pure screws at a certain distance from the
  source and, owing to their high Peierls stress, control its operation.  However, due to the mutual
  interaction between emitted dislocations the group consisting of both non-screw and screw
  dislocations can move at an applied stress that is about a factor of two to three lower than the
  stress needed for the glide of individual screw dislocations.
\end{abstract}

\pacs{}

\keywords{Peierls stress; screw dislocations; mixed dislocations; interactions; Frank-Read sources}

\maketitle


\section{Introduction}

It has been firmly established by many experimental and theoretical studies performed in the last
forty years that the plastic behavior of body-centered-cubic (BCC) metals is controlled by
$1/2\langle{111}\rangle$ screw dislocations the cores of which are non-planar (for reviews see
\cite{kubin:82, christian:83, duesbery:89, vitek:92, seeger:95, pichl:02, duesbery:02}). However,
the only direct experimental observation that suggests such core spreading is the high-resolution
transmission electron microscopic (TEM) study of Sigle \cite{sigle:99} while the primary source of
our understanding of the dislocation core structure and related atomic-level aspects of the glide of
$1/2\langle{111}\rangle$ screw dislocations is computer simulation.  Such calculations have been
made using a broad variety of descriptions of interatomic forces, ranging from pair-potentials
\cite{vitek:70, basinski:71, duesbery:73} to density functional theory (DFT) based calculations
\cite{woodward:01, woodward:02, frederiksen:03} and studies employing other quantum mechanics based
methods \cite{xu:96,xu:98,mrovec:04}.  

The vast majority of atomistic studies of the core structure and glide of $1/2\langle 111 \rangle$
screw dislocations in BCC metals were carried out using molecular statics techniques and thus they
correspond to 0~K.  A problem encountered universally in all the calculations of the critical
resolved shear stress (CRSS), i.e. the Peierls stress, at which the screw dislocation starts to
glide, is that it is by a factor of two to three larger than the CRSS obtained by extrapolating
experimental measurements of the yield and flow stresses to 0~K.  The following are a few examples.
Basinski et al. \cite{basinski:81} measured the flow stress of potassium in the temperature range
1.5~K to 30~K and extrapolated to 0~K to get $0.002\mu$ to $0.003\mu$ where
$\mu=(C_{11}-C_{12}+C_{44})/3$ is the $\langle{111}\rangle\{110\}$ shear modulus and $C_{11}$,
$C_{12}$, $C_{44}$ are elastic constants.  Similar values were found by Pichl and Krystian
\cite{pichl:97b}.  The values of the CRSS when the maximum resolved shear stress plane (MRSSP) is a
$\{110\}$ plane, calculated using a pair potential derived on the basis of the theory of weak
pseudopotentials \cite{dagens:75}, is $0.007\mu$ to $0.009\mu$ \cite{basinski:81}.  More recently,
Woodward and Rao \cite{woodward:02} calculated the CRSS in molybdenum using the many-body potentials
derived from the generalized pseudopotential theory \cite{moriarty:90} and a DFT based method.  When
the MRSSP is a $\{110\}$ plane, they found the CRSS to be between $0.018\mu$ and $0.020\mu$.  A
similar value of the CRSS, $0.019\mu$, was found in calculations employing the tight-binding based
bond-order potential for molybdenum \cite{mrovec:04, groger:05}. Experimental measurements of
Hollang et al. \cite{hollang:01}, extrapolated to 0~K, give for the CRSS in molybdenum $0.006\mu$.
A similar problem was encountered by Wen and Ngan \cite{wen:00} who used the Embedded Atom Method
(EAM) potential for iron and the Nudged Elastic Band method to analyze the activation enthalpies for
kink-pair nucleation on screw dislocations.  The calculated yield stress at 0~K was about $0.013\mu$
while the experimental values, reported by Aono et al. \cite{aono:81} are $0.005\mu$ to $0.006\mu$.
This ubiquitous higher value of the calculated CRSS, found independently of atomic interactions,
suggests that the origin of this discrepancy cannot be sought on the atomic scale of the motion of
individual dislocations but rather on mesoscopic scale where a large number of elastically
interacting dislocations glide at the same time.  In this context it should be noticed that the only
atomistic simulation that predicts yield stress close to that measured experimentally considered
a planar dislocation network of $1/2[111]$ and $1/2[\bar{1}\bar{1}1]$ screw dislocations with
$[001]$ screw junctions \cite{bulatov:02}.  Such a network moved in the $(\bar{1}10)$ plane at the
stress about $50\%$ lower than the Peierls stress of an isolated screw dislocation.

In-situ TEM observations of dislocation sources in BCC transition metals showed that in thin foils
straight screw dislocations formed near the source and moved very slowly as a group \cite{vesely:68,
louchet:75, matsui:76, takeuchi:77, louchet:79, vesely:02, louchet:web}.  Hence they fully control
the rate at which the source produces dislocations.  In the foils used in TEM the applied stresses
are very low but a similar control of the sources by sessile screw dislocations can be expected in
the bulk at stresses leading to the macroscopic yielding. However, at higher stresses dislocations
move faster and do not become pure screws immediately after leaving the source but at a distance
from the source.  Indeed, even in situ observations at higher stresses do not show straight screw
dislocations emanating directly from the sources \cite{vesely:priv06}.

In this paper, we propose a mesoscopic model involving a Frank-Read type source \cite{hirth:82}
emitting dislocations of generally mixed character that become pure screw dislocations at a distance
from the source and, owing to their high Peierls stress, control its operation.  However, there
are a number of non-screw dislocations between the screws and the source, which can move easily.
These dislocations exert a stress on the screw dislocations and this stress, together with the
applied stress, act on the screw dislocations by the force equal to that needed to overcome the
Peierls stress. Screw dislocations can then move at an applied CRSS that is about a factor of two to
three lower than the CRSS needed for the glide of individual screw dislocations.


\section{Model of dislocation nucleation and motion}

\begin{figure}[!htb]
  \centering
  \includegraphics[width=5cm]{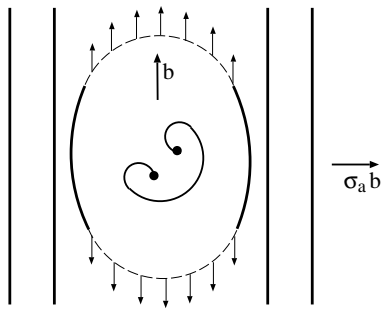}
  \caption{Schematic operation of a dislocation source in bcc metals.  The curved non-screw segments
    migrate away, leaving behind a new pair of screw dislocations.}
  \label{fig_FRsource}
\end{figure}

Let us consider a Frank-Read source (see e.g. \cite{hirth:82}) that produces dislocations in a BCC
metal.  It emits, as always, dislocation loops that have a mixed character and expand easily away
from the source since their Peierls stress is low.  However, at a certain distance from the source,
a significant part of the expanding loop attains the screw orientation and becomes much more
difficult to move owing to the very high Peierls stress of pure screws.  The rest of the loop,
having a mixed character, continues to expand which leads to further extension of the screw segments.
As a result, the source becomes surrounded by arrays of slowly moving screw dislocations, as
depicted schematically in \reffig{fig_FRsource}.  Further operation of the source is hindered by
their back stress and effectively controlled by the ability of the screw dislocations to glide.
 
The operation of the source is driven by the applied stress, $\sigma_a$, which acts by the
Peach-Koehler force \cite{hirth:82} $\sigma_a b$ (per unit length) on the dislocation that bows out.  This
dislocation obviously has a mixed character.  Let us consider now that there are $N_s$ screw
dislocations at distances $x_i$ from the source and $N_m$ dislocations, generally of mixed character,
positioned between the source and the screw dislocations.  We approximate the latter as straight
lines of the same orientation as the screws, positioned at distances $y_k$ from the source, but with
a negligible Peierls stress compared to that of the screws.  In the framework of the
isotropic elastic theory of dislocations the condition for the source to operate is then 
\begin{equation}
  \sigma_a b \geq \frac{\tau}{R} + \frac{\mu b^2}{2\pi\alpha} \sum_{i=1}^{N_s} \frac{1}{x_i} +
  \frac{\mu b^2}{2\pi\beta} \sum_{k=1}^{N_m} \frac{1}{y_k} \ ,
  \label{eq_source}
\end{equation}
where $\tau$ is the line tension of the emitted dislocations, $b$ their Burgers vector, $R$ the
half-length of the source, $\mu$ the shear modulus, and $\alpha$, $\beta$ constants of the order of
unity.  The first term is the force arising from the line tension that pulls the dislocation back
and the second and third terms are forces produced by the stress fields of screw and non-screw
dislocations, respectively, present ahead and/or behind the source.  In the following we neglect the
interaction between dislocations ahead and behind the source as they are far apart.  Moreover, the
dislocation sources are frequently single-ended (see e.g. \cite{hirth:82}).  Hence we analyze only
dislocations ahead of the source, i.e. those towards which the source bows out.

It should be noted here that the screw dislocations in the array ahead of the source are not pressed
against any obstacle and thus they do not form a pile-up.  Within the approximations defined above,
the $i$th screw dislocation will move provided
\begin{equation}
  \sigma_a + \frac{\mu b}{2\pi} \mathop{\sum_{j=1}^{N_s}}_{j\not=i} \frac{1}{x_i-x_j} + 
  \frac{\mu b}{2\pi\alpha} \sum_{k=1}^{N_m} \frac{1}{x_i-y_k} + \frac{\mu b}{2\pi\alpha} \frac{1}{x_i} \geq
  \sigma_P \ ,
  \label{eq_screw}
\end{equation}
where $\sigma_P$ is the Peierls stress of screw dislocations.  The second and third terms are
stresses arising from screw and non-screw dislocations, respectively, and the fourth term is the
stress arising from the dislocation associated with the source that is also treated as a straight
line of the same type as all the other mixed dislocations.  Since the Peierls stress of non-screw
dislocations is negligible, the $l$th non-screw dislocation can move provided
\begin{equation}
  \sigma_a + \frac{\mu b}{2\pi\alpha} \sum_{j=1}^{N_s} \frac{1}{y_l-x_j} + 
  \frac{\mu b}{2\pi\beta} \mathop{\sum_{k=1}^{N_m}}_{k\not=l} \frac{1}{y_l-y_k} + 
  \frac{\mu b}{2\pi\beta} \frac{1}{y_l} \geq 0 \ .
  \label{eq_mixed}
\end{equation}
The meanings of individual terms are analogous to those in equation \refeq{eq_screw}.

Now, the question asked is how large stress, $\sigma_a$, needs to be applied so that the screw
dislocations can move so far away from the source that they either reach a surface or encounter
dislocations of opposite sign from another source and annihilate.  In both cases the source then
keeps producing new dislocations indefinitely.  In the former case these dislocations keep vanishing
at the surface and the latter case leads to the propagation of slip through the sample. In order to
investigate the problem formulated above, we performed the following self-consistent simulations for
certain fixed values of the Peierls stress, $\sigma_P$, and applied stress $\sigma_a$.  First, we
choose a half-length of the source, $R$, and a distance from the source, $y_{max}$, beyond which the
expanding loop always attains the screw character.  The first mixed dislocation emitted by the
source becomes screw when reaching the distance $y_{max}$ and then moves to a distance $x_1$,
determined by equation \refeq{eq_screw}.  Provided that the source can operate, i.e. the inequality
\refeq{eq_source} is satisfied, another dislocation is emitted from the source.  The position of
this dislocation is determined by equation \refeq{eq_mixed} if it does not reach $y_{max}$ and by
equation \refeq{eq_screw} if it does.  Subsequently, the position of the first dislocation, $x_1$,
is updated to satisfy equation \refeq{eq_screw}, which allows also the second dislocation to move.
In this way a new position of the first dislocation, $x_1$, and the position of the second
dislocation, either $y_1$ if smaller than $y_{max}$ or $x_2$ if larger than $y_{max}$, are found
self-consistently.  This self-consistent process is then repeated for the third, fourth, etc.,
dislocations until the source cannot emit a new dislocation, i.e. when inequality \refeq{eq_source}
is no longer satisfied.  The result of this calculation is the number of screw dislocations, $N_s$,
and mixed dislocations, $N_m$, as well as their positions ahead of the source, when the source
becomes blocked by the back-stress from all the emitted dislocations.  The first screw dislocation
is then at a position $x_1=x_{max}$ and further operation of the source can proceed only if this
screw dislocation is removed, as argued above.  The source can then continue operating in a
steady-state manner, producing a large number of dislocations that mediate the macroscopic plastic
flow.


\section{Results}

In the following numerical simulations the applied stress, $\sigma_a$, has been set equal to
$0.3\sigma_P$ and $0.5\sigma_P$, respectively, in order to investigate whether the source can
operate at stress levels corresponding to experimental yield stresses extrapolated to 0~K, as
discussed in the Introduction.  Three values of the Peierls stress, $\sigma_P$, have been considered
that fall into the range found in atomistic studies of transition metals \cite{woodward:02,
mrovec:04, groger:05, wen:00}, namely $0.01\mu$, $0.02\mu$, and $0.03\mu$.  Three different
positions at which mixed dislocations transform into screw ones have been considered, namely
$y_{max}/b = 500, 1000,$ and 2000.  The dependence on the size of the source, $R$, was also
investigated.  However, this dependence is very weak since $R$ enters only through the line tension
term in \refeq{eq_source} and this is always small compared to the terms arising from the
back-stress of emitted dislocations.  Hence, without the loss of generality, we set $R = y_{max}$.
The values of parameters $\alpha$ and $\beta$, entering equations
(\ref{eq_source}) to (\ref{eq_mixed}) have all been set to one and the usual approximation for the
line tension, $\tau=\mu b^2/2$ \cite{hirth:82}, adopted.

\begin{figure}[!htb] 
  \centering
  \includegraphics[width=8cm]{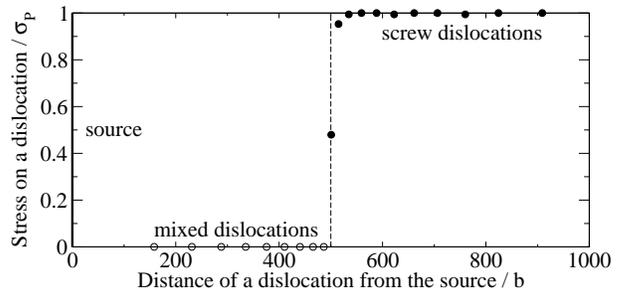}
  \caption{Positions of mixed (open circles) and screw (full circles) dislocations and the
    corresponding stresses when the source is blocked by the back-stress for the case
    $\sigma_a/\sigma_P=0.5$, $\sigma_P/\mu=0.02$, and $y_{max}/b=500$.}
  \label{fig_dstresses}
\end{figure}

Results of such simulation are presented in detail for $\sigma_a = 0.5\sigma_P$, $\sigma_P=0.02\mu$
and $y_{max}=500b$ in \reffig{fig_dstresses}, where positions of the dislocations ahead of the
source and stresses acting on them are shown. In this case $x_{max}/y_{max}=1.8$. It should be noted
that the stress exerted on the majority of screw dislocations is practically equal to their Peierls
stress.  The distances $x_{max}$ found for the above-mentioned two values of $\sigma_a$, three
values of $\sigma_P$ and three values of $y_{max}$, are summarized in \reftab{tab_xmax}.  

\begin{table}[!htb]
  \tabcolsep=5pt
  \begin{tabular}{|c|c|c|c|c|}  
    \hline
    \multicolumn{2}{|c|}{} & \multicolumn{3}{|c|}{$y_{max}/b$} \\ \cline{3-5}
    \multicolumn{2}{|c|}{} & 500 & 1000 & 2000 \\
    \hline
    \multicolumn{5}{|c|}{$\sigma_a/\sigma_P=0.3$} \\
    \hline
                      & $\sigma_P/\mu=0.01$ & 1.0 & 1.1 & 1.2 \\ \cline{2-5}
    $x_{max}/y_{max}$ & $\sigma_P/\mu=0.02$ & 1.2 & 1.3 & 1.3 \\ \cline{2-5}
                      & $\sigma_P/\mu=0.03$ & 1.2 & 1.2 & 1.2 \\ \cline{2-5}
    \hline
    \multicolumn{5}{|c|}{$\sigma_a/\sigma_P=0.5$} \\
    \hline
                      & $\sigma_P/\mu=0.01$ & 1.6 & 1.8 & 2.0 \\ \cline{2-5}
    $x_{max}/y_{max}$ & $\sigma_P/\mu=0.02$ & 1.8 & 2.0 & 2.0 \\ \cline{2-5}
                      & $\sigma_P/\mu=0.03$ & 1.9 & 2.0 & 2.0 \\ \cline{2-5}
    \hline
  \end{tabular}
  \caption{The distance which the leading screw dislocation advances from the source, $x_{max}$, as
    a function of the distance $y_{max}$ from the source at which dislocations become screw, for the
    applied stress $\sigma_a/\sigma_P$ and the Peierls stress of the screw dislocations
    $\sigma_P/\mu$.}
  \label{tab_xmax}
\end{table}

These results suggest that, for a given applied stress, the ratio $x_{max}/y_{max}$ is almost
constant, independent of $y_{max}$, and only weakly dependent on the magnitude of the Peierls stress
$\sigma_P$.  At $\sigma_a/\sigma_P=0.3$, most of the dislocations are mixed and $x_{max}/y_{max}
\approx 1.3$.  With increasing stress, more emitted dislocations become screw and, at
$\sigma_a/\sigma_P=0.5$, $x_{max}/y_{max} \approx 2$, which implies that the numbers of mixed and
screw dislocations ahead of the source are very similar.  Very importantly, the stress exerted on
most of the screw dislocations is practically equal to their Peierls stress, see
\reffig{fig_dstresses}.
	 

\section{Conclusion}

The distinguishing characteristic of the model presented in this paper is that it does not consider
the glide of a single screw dislocation but movement of a large group of dislocations produced by a
Frank-Read type source.  In general, this source produces dislocation loops of mixed character that
transform into pure screws at a distance $y_{max}$ from the source.  Hence, the group of
dislocations consists of screw dislocations at distances larger than $y_{max}$ and non-screw
dislocations near the source.  It is then the combination of the applied stress with the stress
produced by the dislocations in this group that acts on the screw dislocations and is practically
equal to their Peierls stress.  However, after emitting a certain number of dislocations the source
becomes blocked by their back-stress and, at this point, the leading screw dislocation reaches the
distance $x_{max}$ from the source. Nonetheless, the source can continue operating if a dislocation
of opposite sign, originating from another source, annihilates the leading screw dislocation.  This
requires an average separation of sources about $2x_{max}$.  Since the pinning points of
the sources for a given slip system are produced by intersections with dislocations in other slip
systems, their separation is related to the dislocation density in these systems.  For example, in a
deformed molybdenum crystal this density is of the order of $10^{12}\,{\rm m}^{-2}$ \cite{kaspar:00}
which implies separation of dislocations between $3000b$ and $4000b$, for the lattice parameter of
Mo equal to $3.15\,{\rm \AA}$.  These values are in the range of $2x_{max}$ for applied stresses
that are between one-third and one-half of the atomistically calculated Peierls stress for the
sources of the size compatible with the above-mentioned density of dislocations.

The implication of the present study is that the values of the Peierls stress of screw dislocations
in BCC metals found in atomistic studies cannot be compared directly with the yield stress obtained
by extrapolating experimental measurements to 0~K.  The experiments do not determine the stress
needed for the glide of individual screw dislocations but, instead, the stress needed for the
operation of sources that are hindered by the sessile screw dislocations.  These sources can operate
at stresses lower than the Peierls stress owing to the collective motion of screw and mixed
dislocations produced by them, as described in this paper.  Consequently, the discrepancy between
the calculated Peierls stress and the measured yield stress is not a consequence of the inadequacy
of the description of atomic interactions, which has often been raised as a possible explanation,
but incorrectness of their direct comparison.

\vskip1em

\begin{acknowledgments}
  The authors would like to thank Drs. F. Louchet, L. Kubin and D. Vesely for stimulating
  discussions on experimental observations of screw dislocations in BCC metals. This research was
  supported by the U.S. Department of Energy, BES Grant no. DE-PG02-98ER45702.
\end{acknowledgments}

\bibliography{bibliography}

\end{document}